\documentclass[traditabstract]{aa}
\usepackage{txfonts}
\usepackage{graphicx}
\usepackage{graphicx}
\usepackage{natbib}
\bibpunct{(}{)}{;}{a}{}{,}
\usepackage{longtable}

\begin{document}

\title{A large-scale galaxy structure at $z = 2.02$ \\ associated with the radio galaxy MRC~0156-252\thanks{Based on 
observations obtained at the European Southern Observatory using the Very Large Telescope on Cerro 
Paranal through ESO programme 090.A-0734 (P.I. A. Galametz)}}

\titlerunning{A large-scale structure at $z = 2.02$ around MRC~0156-252}

\author{Audrey Galametz\inst{1},
Daniel Stern\inst{2},
Laura Pentericci\inst{1},
Carlos De Breuck\inst{3},
Joel Vernet\inst{3},
Dominika Wylezalek\inst{3},
Rene Fassbender\inst{1,5},
Nina Hatch\inst{4},
Jaron Kurk\inst{5},
Roderik Overzier\inst{6,7},
Alessandro Rettura\inst{8},
Nick Seymour\inst{9}}

\institute{INAF - Osservatorio di Roma, Via Frascati 33, I-00040, Monteporzio, Italy [e-mail: audrey.galametz@oa-roma.inaf.it]
\and {Jet Propulsion Laboratory, California Institute of Technology, 4800 Oak Grove Dr., Pasadena, CA 91109, USA}
\and {European Southern Observatory, Karl-Schwarzschild-Strasse 2, D-85748 Garching, Germany}
\and {University of Nottingham, School of Physics and Astronomy, Nottingham NG7 2RD}
\and {Max-Planck-Institut f\"ur extraterrestrische Physik (MPE), Postfach 1312, Giessenbachstr.,  85741 Garching, Germany}
\and {Observat\'orio Nacional, Rua Jos\'e Cristino, 77. CEP 20921-400, S\~ao Crist\'ov\~ao, Rio de Janeiro-RJ, Brazil}
\and {Department of Astronomy, The University of Texas at Austin, 2515 Speedway, Stop C1400, Austin, TX 78712, USA}
\and {Department of Astrophysics, California Institute of Technology, MS 249-17, Pasadena, CA 91125, USA}
\and {CSIRO Astronomy \& Space Science, P.O. Box 76, Epping, NSW 1710, Australia}
}

\abstract{We present the spectroscopic confirmation of a structure of galaxies surrounding the radio galaxy MRC~0156-252 at $z = 2.02$. 
The structure was initially discovered as an overdensity of both near-infrared selected $z > 1.6$ and mid-infrared selected 
$z > 1.2$ galaxy candidates. We used the VLT/FORS2 multi-object spectrograph to target $\sim$$80$ high-redshift galaxy candidates, and 
obtain robust spectroscopic redshifts for more than half the targets. The majority of the confirmed sources are star-forming galaxies at $z > 1.5$. 
In addition to the radio galaxy, two of its close-by companions ($< 6\arcsec$) also show AGN signatures. Ten sources, including 
the radio galaxy, lie within $\mid z - 2.020 \mid$ $< 0.015$ (i.e.,~velocity offsets $< 1500$~km~s$^{-1}$) and within projected $2$~Mpc 
comoving of the radio galaxy. Additional evidence suggests not only that the galaxy structure associated with MRC~0156-252 is a forming galaxy 
cluster but also that this structure is most probably embedded in a larger scale structure.}

\keywords{Galaxies: clusters: general - Galaxies: clusters: individual: MRC~0156-252 - large scale structure of the Universe - 
Galaxies: individual: MRC~0156-252}

%\titlerunning{A large-scale structure at $z = 2.02$ around MRC~0156-252}
\maketitle

\section{Introduction}

Galaxy clusters are the densest, most massive gravitationally bound regions, emerging from the highest fluctuation peaks 
in the matter distribution of the early universe. They therefore provide crucial constraints on cosmological models 
\citep[e.g.,][]{Jimenez2009} as well as on scenarios of galaxy formation. As large reservoirs of galaxies at a specific 
cosmic time, they also provide unique laboratories to investigate galaxy evolution.
%; studies of galaxies in both 
%clusters and sparser galaxy structures have in particular helped investigate the possible dependance and 
%evolution of galaxy properties with environment.

Decades of studies have drawn a relatively accurate picture of the state and physical processes occurring within 
galaxy clusters at low to intermediate redshift. Clusters possess a core dominated by massive, early-type 
galaxies forming a tight red sequence, up to at least $z \sim 1.5$ \citep{Lidman2008, Nastasi2011}. A primary challenge 
for this field is to push cluster discoveries to higher redshift in order to investigate the 
formation epoch and physical processes that lead to the present-day structures. 

However, until recently, our knowledge of high-redshift galaxy clusters was relatively sparse with only a handful of 
confirmed systems at $z > 1.5$. Discoveries of galaxy clusters at higher redshift were mainly restricted by the limitation of 
classical cluster search methods such as the detection of extended X-ray emission from the intracluster medium 
\citep[e.g.,][]{Stanford2006, Fassbender2011} which is limited by the diminishing surface brightness, the red-sequence  
techniques shifted to the near-infrared at $z > 1.5$, or the promising Sunyaev-ZelÕdovich (SZ) cluster surveys that are, to date, 
still restricted to $z \sim 1.5$ \citep[e.g.,][]{Vanderlinde2010, Foley2011, Stalder2013, Planck2013, Bayliss2013}. 

Over the last decade, two alternative techniques have proven efficient at finding high-redshift galaxy structures. 
On one hand, studies have used the great sensitivity of the {\it Spitzer} telescope and in particular its Infrared Array Camera 
\citep[IRAC;][]{Fazio2004A} to identify $z > 1.2$ galaxy clusters. The IRAC colour $[3.6] - [4.5]$ is very efficient at selecting 
high-redshift galaxies ($z > 1.2$) and therefore at isolating high-redshift clusters \citep{Eisenhardt2008, Papovich2008, 
Galametz2012, Wylezalek2013}. A large number of confirmed high-redshift ($z > 1.5$) galaxy (proto-)clusters were found 
and confirmed through spectroscopic follow-up of overdensities of IRAC red sources --- e.g.,~SXDF-XCLJ0218-0510 
at $z = 1.62$ \citep{Papovich2010, Tanaka2010}, IDCS J1426+3508 at $z = 1.75$ \citep{Stanford2012} and 
CL~J1449-0856 at $z = 2.0$ \citep{Gobat2013}. These galaxy (proto-)clusters were, however, discovered through 
field surveys, i.e.,~serendipitously. Finding larger samples of high-redshift galaxy clusters would require field surveys 
even wider than the several tens of square-degrees which are the current state-of-the-art. 

On the other hand, targeted searches in the surroundings of powerful high-redshift radio galaxies have proven 
quite effective. Radio galaxies are amongst the most massive sources in the Universe with masses of M $> 10^{11}$ 
M$_{\odot}$ \citep{Seymour2007} and preferentially inhabit dense environments. The first searches of 
high-redshift galaxy structures in the environments of radio galaxies typically focussed at $2 < z < 5$. They 
made use of narrow-band imaging to detect emission line objects such as Ly$\alpha$ and H$\alpha$ emitters 
\citep[e.g.,][]{Pentericci2000, Kurk2004B, Venemans2004, Venemans2007, Hayashi2010, Koyama2013} 
that are usually not very massive \citep[a few $10^8$ M$_{\odot}$;][]{Overzier2008} and likely to be young 
star-forming galaxies. An overdensity of star-forming galaxies at $z = 2.3 \pm 0.4$ was also reported in the field 
of the bright $z = 2.72$ QSO HS1700+643 \citep{Steidel2005}. 

Complementary studies have attempted to reveal the massive evolved population in these galaxy structures 
at $z \ge 2$ \citep[e.g.,][]{Kajisawa2006, Tanaka2007, Zirm2008, Galametz2009B, Galametz2010A} in order to 
determine the epoch of appearance of the first red sequence galaxies. However, most of these confirmed 
galaxy structures at $z \ge 2$ are still in the act of collapsing Ñ i.e. they are likely proto-clusters which are   
are not yet part of an evolved and quasi-virialized dark matter halo. Although there has been evidence of a mixed 
population of both star-forming and quiescent galaxies in protoclusters at $z \sim 2$ with the latest formed between 
$z_f \sim 3$ and $4$ \citep{Tanaka2013}, attempts at confirming red sequence objects in proto-clusters at $z \sim 2$ 
have been challenging with even the reddest galaxies showing on-going star-formation (Doherty et al. 2010). 
However, significant star formation activity has also been observed in the cores of massive clusters at $z \sim 1.5$ 
and beyond \citep[e.g.][]{Tran2010, Hayashi2010, Fassbender2011, Bayliss2013, Lotz2013}. The cosmic epoch at $z > 1.5$ 
is therefore crucial for the development of the dominant red and dead cluster galaxy population observed  in lower 
redshift cluster environments. In this respect, detailed studies of galaxy population properties in proto-clusters at $z \ge 2$ 
can provide important insights on the early evolution, environmental effects, and physical processes that drive 
galaxy evolution in dense environments.

Recent programs have intended to combine the two cluster searching techniques mentioned above to systematically 
look for (proto)clusters in the fields of radio galaxies using IRAC $3.6$ and $4.5\mu$m data: e.g.,~the {\it Spitzer} 
High-z Radio Galaxy survey \citep[SHzRG;][]{Galametz2012} and its extension (both in sample size and to type 1 
AGN), the Clusters Around Radio-Loud AGN survey \citep[CARLA;][]{Wylezalek2013}.
 
In this paper, we present spectroscopic observations of a galaxy structure found in the field of the radio galaxy 
MRC~0156-252 ($z = 2.026$) that was part of the SHzRG sample and found to lie in an overdense region of both 
IRAC-selected galaxies \citep[][and Section 3]{Galametz2012} and near-infrared-selected high-redshift galaxy candidates 
\citep[][and Section 2]{Galametz2010B}. 

We follow-up the galaxy structure candidate with the optical multi-object FOcal Reducer and low dispersion Spectrograph 
(FORS2) at VLT. Sections 2 and 3 provide information on the radio galaxy itself and the candidates
selection, respectively. 
%Past literature on the radio galaxy and its associated galaxy structure is summarized in Sections 2 and 3, respectively. 
Section 4 details the observations, data reduction and spectroscopic redshift assignments. Section 5 reports on the 
close-by companions of MRC~0156-252. Section 6 presents the galaxy structure found at $z = 2.020$ 
and a comparison of this structure with current known galaxy clusters and protoclusters at similar cosmic time.
Section 7 summarizes our results.

Throughout the paper, we assume a $\Lambda$CDM cosmology with $H_0 = 70$ km s$^{-1}$ Mpc$^{-1}$, 
$\Omega_m = 0.3$ and $\Omega_{\Lambda} = 0.7$. All magnitudes are expressed in the AB photometric system.% unless stated otherwise. 

\begin{figure*}
\begin{center} 
\includegraphics[width=7cm,angle=0, bb= 200 70 400 250]{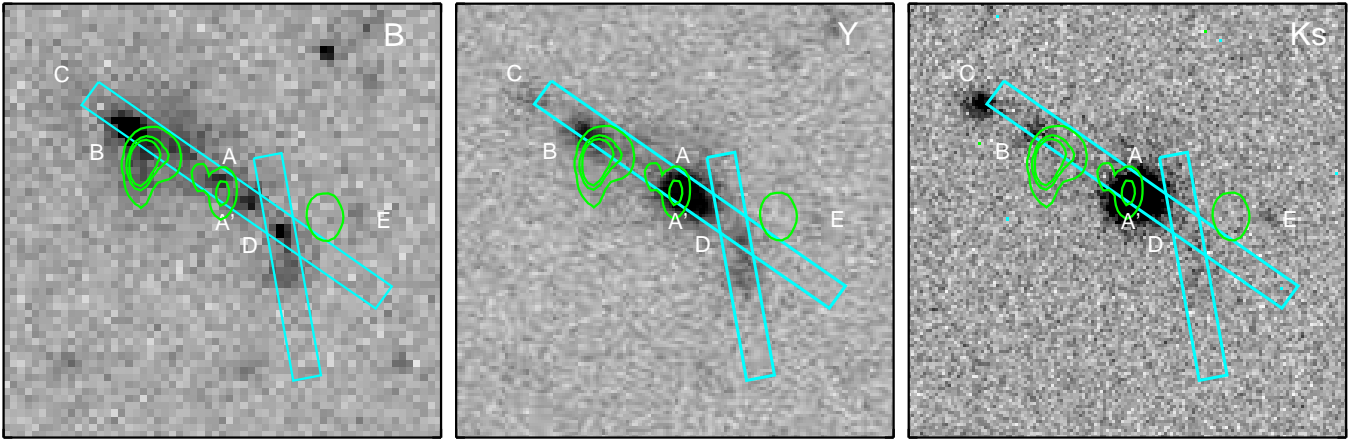}
\end{center}
\caption{The $15\arcsec \times 15\arcsec$ ($\sim$$125$~kpc $\times 125$~kpc at $z \sim 2$) immediate surroundings 
of MRC~0156-252 at optical and near-infrared wavelengths: $B$-band pre-imaging (left panel; this work) and 
$Y$ (middle) and $K_s$ (right) bands (G10). North is up; East is to the left. The two FORS2 slits dedicated to observing the 
different radio galaxy companions are shown in cyan (positions angles of $-55^\circ$ and $-10^\circ$ for the first and 
second mask, respectively). Radio contours from $4.7$~GHz imaging by the VLA are overplotted
in green (Carilli et al.~1997). The radio galaxy and companions are designated by `A', `B', `C', and `D' for consistency with 
the notation of Pentericci et al.~(2001) and G10. We introduce `A$^{\prime}$' for the closest source to A revealed 
by the FORS2 $B$-band image and `E' for a suspected faint companion at the edge of the SW radio lobe.}
\label{zoom}
\end{figure*}

\section{MRC~0156-252}

MRC~0156-252 was first reported in \citet{Large1981} as part of the Molonglo Reference Catalogue 
of Radio Sources. An optical spectrum of the radio galaxy 
%taken with the Schmidt spectrograph on the Cerro Tololo 4 m telescope 
revealed three emission lines that \citet{McCarthy1990} identified as Ly$\alpha$, CIV and HeII at a redshift 
of $z = 2.016$. MRC~0156-252 has narrow Ly$\alpha$ emission 
and shows an extended optical morphology \citep{McCarthy1990} but it is unresolved at near-infrared wavelengths 
\citep{McCarthy1992}. This led \citet{McCarthy1993} to 
conclude that MRC~0156-252 has an old stellar population with some on-going star-formation that could explain 
the extended rest frame UV emission. Based on a later near-infrared spectrum of MRC~0156-252 and the 
study of its broad H$\alpha$ line, \citet{Eales1996} suggested that MRC~0156-252 might actually be a quasar 
largely obscured by dust. They also found that the peak of the H$\alpha$ line was redshifted by 
$\sim$$1000$~km~s$^{-1}$ compared to what expected at $z = 2.016$. Although they note that similar offsets have been observed 
for other high-redshift quasars, we find that the higher quality FORS2 spectroscopy presented in \S4 revises the redshift of 
MRC~0156-252 to $z = 2.0256$, consistent with the position of the H$\alpha$ line from Eales \& Rawlings (1996).
%Although they stated this is consistent with the 
%object being a quasar --- with similar offsets having been observed for other high-redshift quasars --- we propose 
%an alternative explanation. The new optical spectrum obtained as part of the present FORS2 follow-up indeed 
%reveals five strong emission lines corresponding to Ly$\alpha$, CIV, HeII, AlIII and CIII] at a revised redshift 
%of $2.0256$ (see \S6), a result consistent with the position of the H$\alpha$ line from \citet{Eales1996}.

Radio galaxies show a range of morphologies often with clumpy structures \citep{Pentericci1999, Galametz2009B}. 
MRC~0156-256 has, at least, four bright close-by companions previously reported in 
McCarthy et al.~(1990), Pentericci et al.~(2001) and Galametz et al.~(2010a),\nocite{Pentericci2001, McCarthy1990} including 
the brightest source (in the near-infrared) usually identified as the radio source.  Figure~\ref{zoom} shows the $15\arcsec \times 15\arcsec$ 
immediate surroundings of MRC~0156-252 in $B$ (pre-imaging; see \S4.1), $Y$ and 
$K_s$. The majority of the companions are aligned in the direction of the radio axis ($\sim$ NE / SW) and spread 
over less than $11\arcsec$. Source A is the brightest source that coincides with the position of the radio core. 
Source B is located at the very edge of the NE radio lobe and has very blue colours, possibly due to enhanced 
star-formation triggered by the radio jet \citep{Pentericci2001}. A faint companion (E) is detected at the edge of the SW 
radio lobe. Source D, found $\sim$$5\arcsec$ SW of A, also has blue colours. The FORS2 $B$-band image reveals 
that D has a disturbed morphology with a brighter compact `core' and an extension to the south. Source C is located 
$\sim$$6\arcsec$ NE of A and has red near-infrared colours. An additional source (A$^{\prime}$) was also 
revealed by the optical data within $1\farcs5$ of the radio galaxy.
%, although the elongated morphology of the radio galaxy to the south-west at near-infrared wavelengths already 
%suggested the presence of two bright close-by components. 

\section{Identification of a galaxy cluster candidate in the field of MRC~0156-252}

As previously mentioned, several pieces of evidence suggest that MRC~0156-252 is embedded in a 
(proto)cluster at high redshift. 

\begin{figure*}%[!t] 
\begin{center} 
\includegraphics[width=13cm,angle=0]{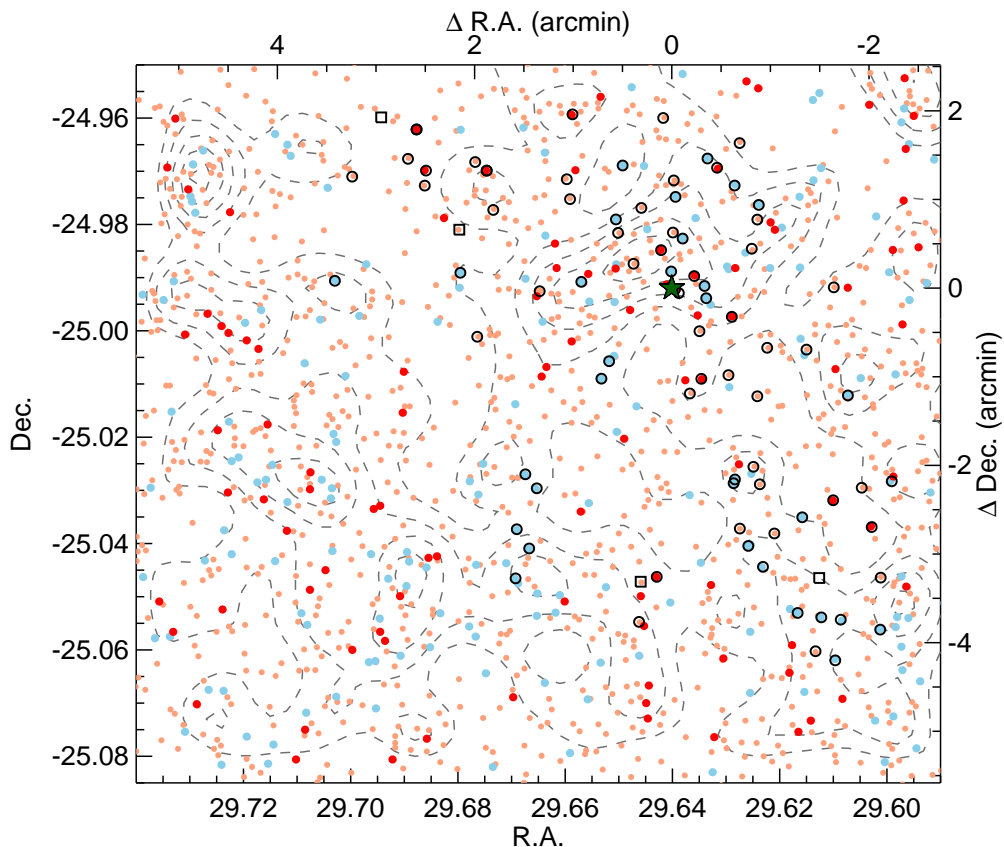}
\end{center}
\caption{Spatial distribution of high-redshift galaxy candidates in the field of MRC~0156-252. Blue and red 
circles show b-$YHK$ and r-$YHK$, respectively ($z > 1.6$ candidates; G10). Pink circles show IRAC-selected sources 
($z > 1.2$ candidates; G12). MRC~0156-252 is shown by the green star. Density distribution of the full $YHK$ + IRAC
selected sample are shown by the dashed gray contours. The spectroscopic targets are highlighted by open black 
circles.  Four bright, low-redshift galaxies chosen for mask alignment are shown by open black squares. We note that
two have colours consistent with IRAC-selected sources which shows the possible contamination of the IRAC selection
by low-redshift galaxies.}
%The dashed ellipse corresponds to the region of diffuse X-ray emission reported by \citet{Overzier2005}.}
\label{radec}
\end{figure*}

MRC~0156-252 was found to lie in an overdensity of near-infrared selected $z > 1.6$ galaxy candidates (Galametz et 
al.~2010a; G10 hereafter)\nocite{Galametz2010B}. G10 introduces a purely near-infrared two-colour $YHK$ selection 
criterion to isolate galaxies at $1.6 < z < 3$ and differentiate between old, red stellar populations and young, blue star-forming 
galaxies, designated as r-$YHK$ and b-$YHK$, respectively. We refer to G10 for details on the colour selection technique. 
At the depth of the near-infrared data, the field of MRC~0156-252 was found to be $\sim3$ times denser than control fields in 
$YHK$-selected galaxies including the GOODS-South field, which is  known to host a galaxy protocluster at $z \sim 1.6$ 
\citep{Castellano2007, Kurk2009}. Half of the r-$YHK$ galaxies have colours consistent with red sequence models at 
$z \sim 2$. G10 therefore concluded that MRC~0156-252 is likely associated with a galaxy (proto)cluster at $z \sim 2$. 
Figure~\ref{radec} shows the spatial distribution of the r-$YHK$ and b-$YHK$ selected sources in the field of MRC~0156-252.

MRC~0156-252 was also found to lie in an overdensity of mid-infrared selected $z > 1.2$ galaxy candidates 
(Galametz et al.~2012; G12 hereafter)\nocite{Galametz2012}. G12 studied the environment of $48$ high-redshift 
($1.2 < z < 3$) radio galaxies --- including MRC~0156-252 --- in order to identify potential associated galaxy clusters. 
G12 made use of a single red IRAC colour cut ($[3.6] - [4.5] > -0.1$) to isolate $z > 1.2$ galaxies (Papovich et al.~2008, 
Wylezalek et al.~2013)\nocite{Papovich2008, Wylezalek2013}. The field of MRC~0156-252 was the ninth densest 
field in red IRAC-selected galaxies, overdense at a $3\sigma$ level compared to average. Its surface density of 
IRAC-selected galaxies was found to be consistent with the CL~J1449+0856 $z \sim 2$ protocluster \citep{Gobat2011}. 
G12 therefore also concluded that MRC~0156-252 is indeed embedded in a high-redshift galaxy (proto)cluster.
The spatial distribution of the red IRAC-selected sources in the field of MRC~0156-252 is also reported in Figure~\ref{radec}.

\section{VLT/FORS2 spectroscopy}

We conducted spectroscopic observations with VLT/FORS2 to confirm the membership of the cluster member candidates 
in the field of MRC~0156-252. 

\subsection{Mask design and targets}

Only limited shallow optical data were available in the field of MRC~0156-252 prior to the FORS2 run from Las Campanas 
\citep[$r$, $I$ and Ly$\alpha$ data;][]{McCarthy1990, Pentericci2001} and {\it HST}/WFPC2 \citep[$F555W$; ][]{Overzier2005}.
To accurately design the masks for the FORS2 spectroscopy, we obtained deep ($3.8$~hr) imaging 
in the $b_{\rm HIGH}$ band ($\lambda = 440$~nm corresponding to $\sim$$145$~nm rest frame at $z = 2.02$; referred to 
as $B$ hereafter). The $B$-band was chosen to match the wavelength range covered by the spectroscopy. The data were 
reduced using standard {\tt IRAF} reduction tools and source extraction was done using 
SExtractor \citep{Bertin1996}.

We designed two slit masks and chose their orientations to optimize the observations of the radio galaxy 
and its neighbors i.e.,~positions angle of $-55^\circ$ for the first mask and $-10^\circ$ for the second. 
Figure~\ref{zoom} shows the two slits that were designed to 
observe the radio galaxy companions. One slit of the first mask was placed along the radio axis to target 
sources A, A$^{\prime}$, B and D.
%The slit was shifted towards the south for an optimal sky subtraction. As a consequence, 
We did not observe source C and chose to elongate the slit towards the south for an optimal sky subtraction.
Source C is barely detected in the $B$-band and thus it is expected to be quite challenging to 
confirm through optical spectroscopy. One slit of mask 2 targeted the 
southern extended companion of the radio galaxy (source D and its extension to the south). The rest of the masks 
were designed as follows: the first mask was preferentially filled with sources that are relatively bright in the 
$B$-band ($B < 23$). The second mask was filled with sources down to a fainter ($B < 24$) magnitude.

In total, we targeted $83$ distinct sources (two r-$YHK$ selected sources were common to both masks): the radio galaxy 
and neighbors ($2$ slits), $12$ r-$YHK$ selected sources (including $6$ IRAC-selected sources), 
$31$ b-$YHK$ selected sources (including $25$ IRAC-selected sources), $33$ red IRAC-selected sources, one 
mid-infrared selected AGN candidate (G12) --- identified by the four IRAC bands AGN colour wedge from \citet{Stern2005} --- and 
four additional bright, nearby sources to check the mask alignment. %The list of targets is provided in Table~\ref{targets}.

\subsection{Observations and data reduction}

The two slit masks were observed on UT 2012 October 20 using the MXU mode of FORS2 with the blue-optimized 
detector and the 300V+10 grism with no order-sorting filter to cover a typical wavelength range of $3300-
6600$\AA~($1100-2200$\AA~at $z = 2$). Mask~1 and Mask~2 (filled with fainter targets) were exposed for 
$3$~hours and $4.4$~hours total, respectively. The sky conditions were clear with an average $0.6''$ seeing 
throughout the night.

We reduced the spectroscopic data using `BOGUS'\footnote[1]{http://www.ucolick.org/\~holden/datareducetext/bogus.html}, 
an {\tt IRAF} script designed for efficient processing of multi-slit
data.  The pipeline includes classical reduction steps such as flat-fielding, cosmic ray rejection, sky subtraction, and 
combination of independent exposures.  The 1D spectra were extracted using {\tt IRAF}. The uncertainties on the
wavelength calibration were estimated from the sky lines to be $\sim$$0.2$\AA.

\subsection{Spectroscopy results}

\begin{figure}%[!t] 
\begin{center} 
\includegraphics[width=8.5cm,angle=0,bb=40 60 550 400]{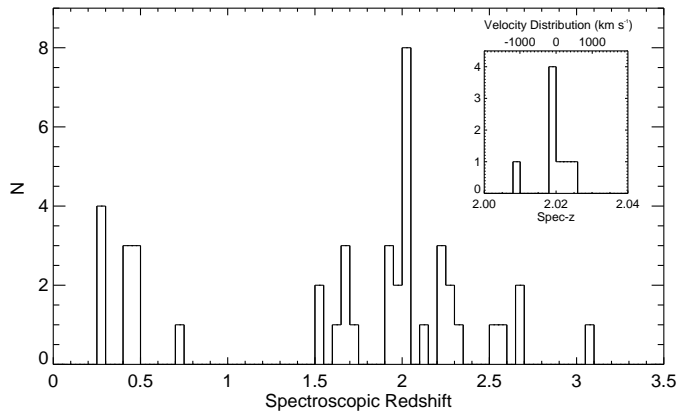}
\end{center}
\caption{Distribution of the available spectroscopic redshifts. The inset panel zooms in on the redshift range $2.00 < z < 2.04$. 
The top axis indicates the corresponding velocity distribution relative to $z = 2.020$.}
\label{dist}
\end{figure}

Spectroscopic redshifts ($z_{\rm spec}$) were assigned to $43$ galaxies. The redshifts were determined using EZ \citep{Garilli2010}.
%by fitting absorption and emission lines with Gaussian profiles using {\tt IRAF}. 
The majority of the derived redshifts are at a high-confidence level with 
at least three strong absorption lines or two strong emission lines. Quality flags `B' were assigned to redshifts with 
weakly detected absorption lines. Redshift uncertainties were derived by summing in quadrature the fitting and wavelength 
calibration uncertainties. Figure~\ref{dist} shows the distribution of our spectroscopic redshifts. Appendix A reports the identification 
number (in the $B$-band) of these $43$ sources with $z_{\rm spec}$, coordinates, target selection technique, spectroscopic 
redshifts and detected absorption/emission lines for the members of the galaxy structure associated with MRC~0156-252 (see 
section 6 and Table~\ref{spectro}) and for foreground/background galaxies (Table~\ref{spectro2}). 

Mainly due to the wavelength coverage probed by the FORS2 configuration, we did not confirm any galaxy in the redshift 
range $1 < z_{\rm spec} < 1.5$. Galaxies with $z_{\rm spec} < 1$ were mostly confirmed thanks to their [OII] emission 
doublet. A large fraction of the galaxies with $z_{\rm spec} > 1.5$ are star-forming galaxies and present 
a clear continuum discontinuity due to the Ly$\alpha$ forest in addition to standard absorption lines (e.g.,~SiII$_{1260}$; 
OI/SiII; CII; SiIV; SiII$_{1526}$; CIV; FeII; AlII; SiII$_{1808}$; AlIII). Four sources have AGN signatures at 
$2.0 < z_{\rm spec} \leq 2.2$ including sources B and D. Due to the adopted spectrograph configuration, 
we should in principle be able to detect Ly$\alpha$ for galaxies at $z > 1.8$, provided they are star-forming and do not 
contain dust. A Ly$\alpha$ emission line was observed in $8$ of the $24$ galaxies confirmed at $z > 1.8$.
% as well as two r-$YHK$-selected sources. whose red near-infrared colours could be a consequence of hosting an AGN.

%The 300V+10 grism was specifically chosen to target the Ly$\alpha$ emission line. Due to the adopted spectrograph 
%configuration, we should in principle be able to detect Ly$\alpha$ for galaxies at $z > 1.8$, provided they are star-forming 
%and do not contain dust. Spectra of $\sim$$40$\% of our galaxies with $z_{\rm spec} > 1.8$ ($32$\% for `normal non-AGN' galaxies) 
%clearly show Ly$\alpha$ emission. 
%({\bf Comparison with literature / presence of Ly$\alpha$ in random high-z galaxy survey?}).

We obtained spectroscopic redshifts for $32$ of the $58$ ($\sim55$\%) targeted IRAC-selected galaxies. $27$ of them 
($\sim$$84$\%) have, as expected from their mid-infrared colours, $z_{\rm spec} > 1.2$ which shows that a simple red IRAC 
colour cut is particularly efficient at identifying high-redshift galaxies. At least one of the lower redshift galaxies 
is an obscured AGN, which is a population known to have red mid-infrared colours \citep[e.g.,][]{Stern2005, Stern2012}, 
and thus are a known contaminant to the simple $[3.6] - [4.5]$ selection. 
%Since the spectroscopic identification of galaxies at low-redshift ($z_{\rm spec} < 1.2$) is generally less challenging 
%than at higher redshift, this $\sim$$80$\% reliability of the IRAC criterion is most probably a lower limit. 
We also derived spectroscopic redshifts for 
$14$ of $31$ ($45$\%) targeted b-$YHK$ galaxies out of which $11$ have a $z_{\rm spec} > 1.6$ and $13$ with 
$z_{\rm spec} > 1.5$. This result demonstrates the efficiency of near-infrared broad-band selection at preferentially isolating 
high-redshift galaxies. To our knowledge, this spectroscopic run is the first to follow-up $YHK$-selected galaxies and 
demonstrates the efficiency of the selection technique. However, it would be highly speculative in the light of this first preliminary 
result to derive estimates of the effectiveness of the $YHK$ selection criterion. We were not able to derive spectroscopic 
redshifts for half of the targets which probably lack strong absorption/emission lines out of their faint continuum. 
We can not dismiss however that they could be lying at a redshift not probed by our FORS configuration.
% or lying at the expected redshift but too faint to derive a reliable spectroscopic redshift.

Only three of the $12$ targeted r-$YHK$ were assigned a spectroscopic redshift, two of which are type-2 AGN. The confirmation
of red galaxies at high redshift is known to be challenging since they are not expected to show strong features, especially at 
optical wavelengths. For example, \citet{Wuyts2009} followed-up a sample of Distant Red Galaxies (DRG; $J - K > 2.3$)
using either deeper observations ($> 8$~hrs) with VLT/FORS2 or with similar exposure times on Keck/LRIS and were able 
to derive spectroscopic redshifts for about $22$\% of their sample. Their $14$ $z > 2$ confirmed DRG all show Ly$\alpha$. 
Similarly, $2$ of our $12$ r-$YHK$ show a $Ly\alpha$ emission line.

\section{The radio galaxy companions}

\begin{figure*}
\begin{center} 
\includegraphics[width=4.8cm,angle=90, bb=200 50 400 800]{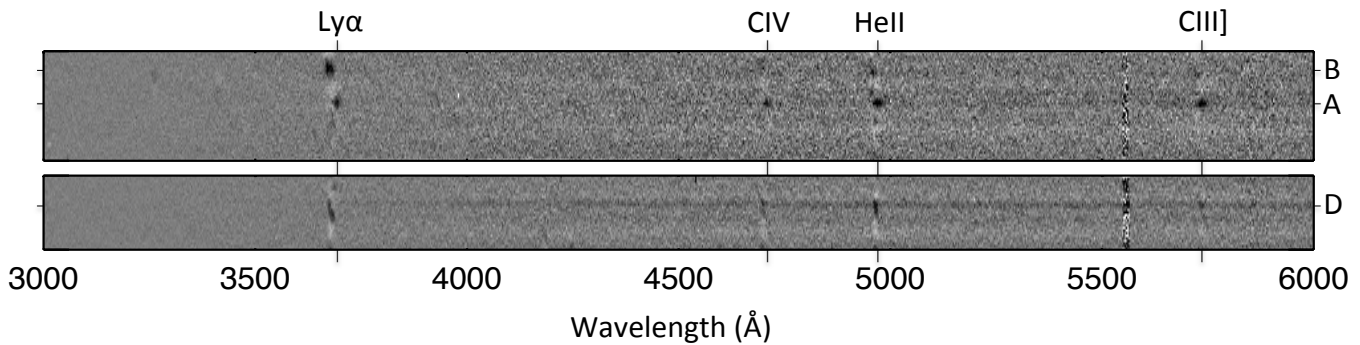} 
\end{center}
\caption{2D spectra of the radio galaxy and close-by neighbours (inverse gray scale) corresponding to the two slits in Fig.~1 with position angles of $-55^\circ$ 
(top; sources A and B) and $-10^\circ$ (bottom; source D). The vertical lines point to the position of the four detected 
emission lines (Ly$\alpha$ $\lambda1216$, CIV $\lambda1550$, HeII $\lambda1640$, CIII] $\lambda1909$) at the redshift of A. 
Continua of sources A, B and D are indicated by the horizontal ticks.}
\label{cut}
\end{figure*}

\begin{figure*}
\begin{center} 
\includegraphics[width=18cm,angle=0]{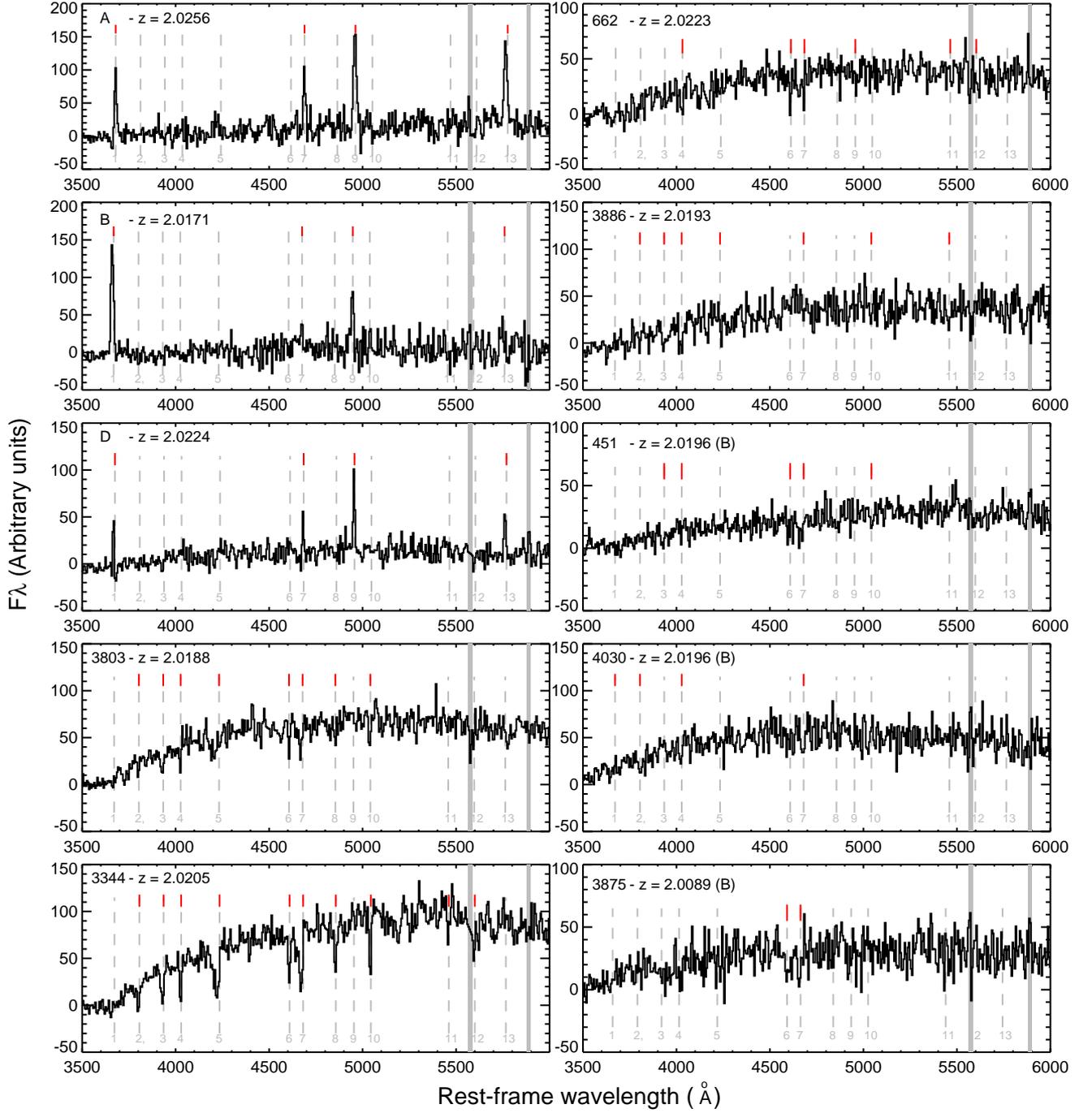} 
\end{center}
\caption{Rest-frame UV spectra obtained with FORS2 for the ten confirmed cluster members of the galaxy structure at $z = 2.02$. 
Identification numbers from the $B$-band catalogue and redshift (followed by `(B)' if it is a low quality redshift) are indicated at the upper left 
corner of each spectrum. The spectra were rebinned by a factor of two. Sky lines are indicated by gray vertical lines. Prominent 
emission and absorption lines expected in this spectral range are shown by the vertical dotted lines: Ly$\alpha$ $\lambda1216$ (1), 
SiII $\lambda1260$ (2), OI/SiII (3; OS, $\lambda1303$), CII $\lambda1334$ (4), SiIV $\lambda1402$ (5), SiII $\lambda1526$ (6), CIV $\lambda1550$ (7), 
FeII $\lambda1608$ (8), HeII $\lambda1640$ (9), AlII $\lambda1670$ (10), SiII $\lambda1808$ (11), AlIII $\lambda1854$ (12) and CIII] $\lambda1909$ (13). 
The absorption/emission lines detected in each individual spectrum are marked by the red ticks and also listed in 
Table~\ref{spectro}.}
%We note that the spectrum of source \#622 also shows an emission line at $\lambda_{\rm rest} \sim 1370$\AA~which could not 
%be identified at the redshift of the source. Since this source has several close-by projected neighbors (see Figure~\ref{snap}), we 
%suggest this line most probably comes from another object along the line-of sight.}
\label{spectra}
\end{figure*}

A redshift was derived for sources A, B and D. Their 2D and 1D spectra are shown in Figures~\ref{cut} and \ref{spectra}
respectively. Sources C and E were not targeted. As far as A$^{\prime}$ is concerned, its proximity to A and the width 
of A's emission lines made it unfeasible to extract a reliable spectroscopic redshift. 

Source A was assigned a redshift of $z = 2.0256 \pm 0.0002$ derived by fitting a Gaussian profile to the HeII emission 
line, a non-resonant line, therefore not affected by absorption. HeII is also the strongest emission line in 
this part of the spectrum. \citet{McCarthy1990} had previously assigned a redshift of $z = 2.016$ through 
optical spectroscopy while \citet{Eales1996} found through near-infrared spectroscopy that the 
H$\alpha$ line was redshifted by $\sim$$1000$~km~s$^{-1}$. Results from \citet{Eales1996} are 
consistent with the revised redshift derived from the FORS2 data. The \citet{McCarthy1990} redshift is however consistent 
with the redshift of companion B ($z = 2.0171 \pm 0.0004$). B is the brightest source in the $B$-band.

Sources A, B and D of MRC~0156-252 show AGN signatures with prominent CIV, HeII and CIII] emission. Their 
clearly distinct continuum and emission lines (see Figure~\ref{cut}) indicate that these sources are separated. The 
detected lines are therefore not a signature of extended halos of the radio galaxy but of individual galaxies, although 
extended CIV and HeII halos are commonly found in radio galaxies \citep{Villar2007}. The emission lines of source D 
show a spatial extension to the south however, indicating that the cloud-like extension to the south of D is associated 
with D and not just an additional independent source along the line of sight.

Several radio galaxies are also known to have companion AGNs. \citet{DeBreuck2010} presented the {\it Spitzer}/IRAC 
follow-up of a sample of 70 radio galaxies and found that at least four have companions within 
$6\arcsec$ with mid-infrared colours consistent with AGN. The spectroscopic follow-up of one of them, 
7C~1756-152 at $z = 1.42$, showed a companion AGN, offset by $6\arcsec$ from the radio galaxy 
\citep{Galametz2010A}. 
%The well-studied Spider Web galaxy (PKS~1138-262) at $z = 2.16$ is also known to have 
%a nearby component with an AGN signature \citep{Kurk2003}. 

These results, including the clumpy morphology of the radio galaxies (evident for MRC~0156-252; see 
Figure~\ref{zoom}), could be suggestive of merger-triggered AGN activity, as predicted by several models 
of galaxy evolution \citep[e.g.,][]{Hopkins2008}. All three AGN in MRC~0156-252 are type-2, or obscured 
AGN whose obscuration could be primarily caused by galaxy-scale gas and dust.  The strong feedback 
associated with these multiple powerful AGN could rapidly expel gas from the merging central galaxy of 
this (proto)cluster, shutting down star formation and leading to the old, red ellipticals that we see associated 
with rich environments in the local universe.

Table~\ref{rgc} lists the redshifts and velocity dispersions of the detected emission lines of sources A, B and D. 
Redshifts were assigned by fitting single Gaussian profiles to the emission lines. Velocity offsets are derived 
relative to the HeII line of source A. The redshifts of A, B and D listed in Table~\ref{spectro} correspond to the 
redshift derived from the HeII line. A large range of velocity shifts is observed between the different companions 
(i.e.,~hundreds of km~s$^{-1}$). Velocity shifts vary from a couple of hundreds to $\sim$$1300$~km~s$^{-1}$, 
a result similar to the range of velocities observed among the different components
of PKS~1138-262 \citep{Kuiper2011}.

\begin{table}
\caption{Emission lines of the radio galaxy and companions}
\centering
\begin{tabular}{l l l l l ll}
Source	&	Line			&	Spec. z				&	Velocity$^{\mathrm{a}}$	\\
			&				&						&	km~s$^{-1}$ \\
\hline
\hline
A			&	Ly$\alpha$	&	$2.0276	\pm	0.0004$	&	$+200$		\\
			&	CIV			&	$2.0288	\pm	0.0007$	&	$+320$		\\
			&	HeII			&	$2.0256	\pm	0.0002$	&	$0$			\\
			&	CIII]			&	$2.0215	\pm	0.0002$	&	$-410$		\\
\hline
B			&	Ly$\alpha$	&	$2.0128	\pm	0.0005$	&	$-1270$		\\
			&	CIV			&	$2.0219	\pm	0.0005$	&	$-370$		\\
			&	HeII			&	$2.0171	\pm	0.0004$	&	$-840$		\\
%			&	CIII]			&	$$		&	$$	\\
\hline
D			&	Ly$\alpha$	&	$2.0183	\pm	0.0001$	&	$-720$		\\	
			&	CIV			&	$2.0237	\pm	0.0002$	&	$-190$		\\
			&	HeII			&	$2.0224	\pm	0.0002$	&	$-320$		\\
			&	CIII]			&	$2.0192	\pm	0.0002$	&	$-640$		\\	
\hline
\end{tabular}
\label{rgc}
\begin{list}{}{}
\item[$^{\mathrm{a}}$] Relative to the HeII emission line of the radio galaxy (source A).
\end{list}
\end{table}

%\section{Discussion}

\section{A galaxy structure at z = 2.02}

\begin{figure}
\begin{center} 
\includegraphics[width=9cm,angle=0]{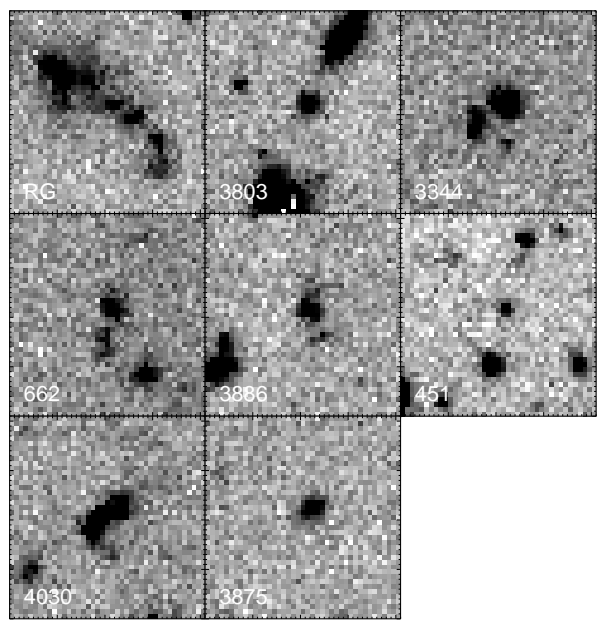}
\end{center}
\caption{$B$-band images of the radio galaxy and companions (first panel) and the seven additional spectroscopically confirmed structure 
members. Images are $10\arcsec $ on a side, with North up and East to the left.}
\label{snap}
\end{figure}

\begin{figure}%[!t] 
\begin{center} 
\includegraphics[width=8.5cm,angle=0]{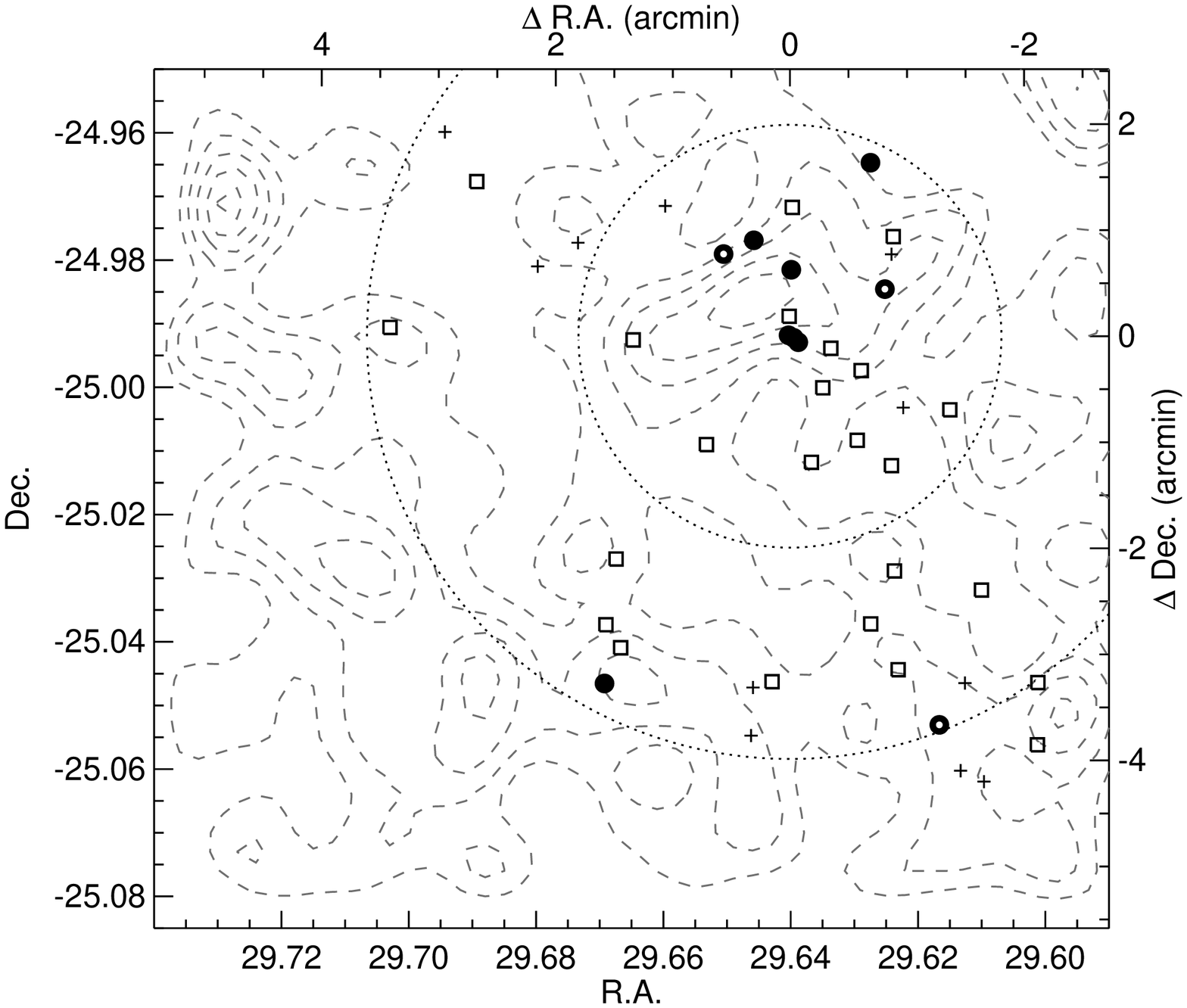}
\end{center}
\caption{Spatial distribution of the spectroscopic redshifts. The ten members of the structure at $z = 2.02$ 
are shown by filled circles (for high-quality redshifts) and open circles (for quality `B' $z_{\rm spec}$). The other 
sources with an assigned spectroscopic redshifts are shown by crosses for low-redshift ($z_{\rm spec} < 1$) galaxies 
and open squares for high-redshift ($z_{\rm spec} > 1.5$) galaxies. 
%No galaxies were confirmed in the redshift range  $1 < z_{\rm spec} < 1.5$. 
Proper distances of $1$ and $2$~Mpc (at $z=2.020$) from the radio galaxy are shown by the dotted lines. The density 
contours from Fig.~\ref{radec} are also included.}
\label{radec2}
\end{figure}

The distribution of spectroscopic redshifts in Figure~\ref{dist} clearly shows a peak at $z \sim 2.02$. 
Ten galaxies, including the radio galaxy, have redshifts within $\mid$$z - 2.020$$\mid$ $< 0.015$, 
corresponding to velocity offsets $< 1500$~km~s$^{-1}$ relative to $z = 2.020$ (nine within 
$600$~km~s$^{-1}$). %Assuming the structure is a virialized galaxy cluster at high-redshift, 
Adopting a biweight estimator \citep{Beers1990} for galaxy clusters with a small number of confirmed 
members, we estimate a line-of-sight velocity dispersion of $\sigma \sim 580$~km~s$^{-1}$, 
or $\sim$$380$~km~s$^{-1}$ when only considering members within $1.5\arcmin$. 
Spectra of the structure members and $B$-band images are shown in 
Figure~\ref{spectra} and Figure~\ref{snap}, respectively. They were all IRAC-selected galaxies; four 
were also selected as b-$YHK$ galaxies. Except for the radio galaxy companions, all spectroscopically 
confirmed structure members are classical star-forming galaxies. 

With a limited number of galaxies confirmed with spectroscopy, it is premature to draw conclusions 
on the nature of the galaxy structure. It could either be an already assembled, relaxed galaxy 
cluster or a protocluster still in formation. We will take a closer look at these possibilities below.

\subsection{A large-scale structure}

Figure~\ref{radec2} shows the spatial distribution of the structure members alongside other sources with a 
spectroscopic redshift from our FORS2 run. Five structure members (not including the radio galaxy 
companions) were found within $1$~Mpc of the radio galaxy. The two other confirmed members were found at a proper 
distance of $\sim$$2$~Mpc south/south-east of MRC~0156-252. Both G10 and G12 
suggested that the galaxy structure associated with MRC~0156-252 may extend to the south east 
with a second overdensity of both b-$YHK$ and IRAC-selected galaxies found $2$~Mpc away from the 
radio galaxy (i.e.,~R.A. $\sim 29.68$; Dec. $\sim -25.05$). This FORS2 spectroscopic follow-up 
of the structure focused on confirming cluster members near the radio galaxy and only a few high-redshift 
galaxy candidates were targeted in that second overdensity. However, source \#662 in that structure was 
confirmed at $z = 2.0223$ which suggests this second overdensity may also be associated with the 
galaxy structure in which MRC~0156-252 is embedded. Since both the near-infrared and IRAC data only cover 
the radio galaxy surroundings and immediate south east region (G10), we do not rule out that the structure 
could also extend beyond our field of view.

%Although the statistic derived is uncertain due to the limited number of confirmed cluster members, the estimated 
%value is consistent with velocity dispersion estimates derived in galaxy clusters at high redshift (e.g.,~Mullis et al.~2005, 
%Hilton et al. ~2007, Eisenhardt et al.~2008, Papovich et al.~2010, Gobat et al.~2011)\nocite{Hilton2007, Eisenhardt2008, 
%Mullis2005}, at intermediate redshift \citep[e.g.,][]{Stanford2001} and for local galaxy clusters such as the Virgo cluster. 

\begin{figure} 
\begin{center} 
\includegraphics[width=8cm,angle=0,bb= 60 60 550 500]{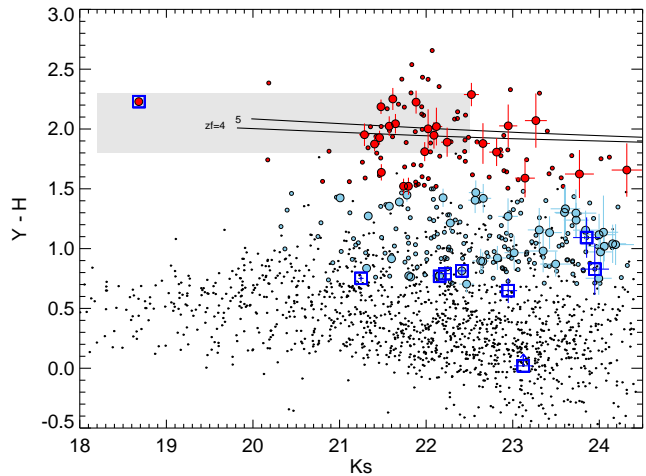}
\end{center}
\caption{Colour-magnitude diagramme $Y - H$ versus $K_s$ (based on G10) of sources detected in $YHK_s$ in the 
environment of MRC~0156-252 (black dots). The r-$YHK$ and b-$YHK$-selected galaxies are shown in red and blue 
circles respectively. Bigger symbols (with error bars) mark sources within $1$~Mpc of the radio galaxy (at $z = 2.02$). 
Red sequence models at $z = 2.02$ for formation redshifts of $4$ and $5$ are shown by the solid lines (G10). 
Spectroscopically confirmed structure members ($\mid$$z_{\rm spec} - 2.02$$\mid$ $< 0.015$) are shown by blue 
squares (with error bars). The radio galaxy is the bright red source with $K_s \sim 18.7$. The gray box designates 
r-$YHK$ galaxies with $K_s < 22.5$ and $1.8 < Y - H < 2.3$ whose density was compared to fields surveys in Section 6.2. 
The FORS2 observations did not permit us to confirm (or deny) redshifts for any of the red sequence member candidates.}
\label{cmd}
\end{figure}

\subsection{A red sequence?}

Figure~\ref{cmd} shows the distribution of sources in the surroundings of MRC~0156-252 in a $Y - H$ versus $K_s$ 
colour-magnitude diagramme (see G10 for details on the $YHK_s$ photometry). The $YHK$-selected galaxies 
i.e.,~galaxy candidates at $z > 1.6$, show the classical bimodal distribution in colours observed for galaxies up to at 
least $z \sim 2$ \citep{Williams2009} with two distinct populations of galaxies with blue and red near-infrared colours, 
isolated by design by a $Y - H \sim 1.5$ colour cut. We note however that the red candidates within $1$~Mpc of the radio 
galaxy preferentially lie on a `sequence', similar to the red sequence observed in virialized galaxy clusters. The mean 
of the red galaxies is $Y - H \sim 1.94$ with a $3\sigma$-clipped scatter of $\sim0.22$, a colour consistent with red 
sequence models at $z = 2$ with $z_f = 4-5$ (G10).

We isolate the bright end of the potential proto-cluster red sequence i.e.,~galaxies with $K_s < 22.5$, $1.8 < Y - H < 2.3$
(see Fig.~\ref{cmd}, gray box; $14$ sources) and compare its density to field surveys. We made use of the public 
multiwavelength catalogue of the CANDELS UDS field \citep{Galametz2013} which contains $\sim 36000$ sources over 
an area of $\sim 200$ square arcmin. The CANDELS UDS field was also imaged by VLT/HAWK-I $YK_s$ bands as well 
as UKIRT/WFCAM $H$-band (UKIDSS DR8), consistent with the VLT/HAWK-I $H$ filter. No filter correction is therefore 
required. The average density of the $K_s < 22.5$ r-$YHK$-selected galaxies with $1.8 < Y - H < 2.3$ derived from the 
UDS is $1320 \pm 160$ per square degree. The density of such sources in the immediate surroundings of MRC~0156-252 
(i.e.,~$< 1$~Mpc $\sim 2\arcmin$ at $z \sim 2$) is $4010 \pm 1070$ per square degree i.e.,~$3$ times higher than in the field.

Apart from the radio galaxy, the brightest red candidates have a $K_s \sim 21$, similar to the bright end limits of the 
suspected red sequences of both CL J1449+0856 \citep[at $z = 2.0$;][]{Gobat2011} and IDCS J1426+3508 
\citep[$z=1.75$;][$F160W \sim 22$]{Stanford2012}. The only two red galaxies confirmed to reside in the Spider 
Web galaxy protocluster also have $K_s \sim 21$ \citep{Doherty2010}. These two galaxies, both confirmed from 
H$\alpha$ emission spectroscopy, are amongst the brightest galaxies in that field in the near-infrared.

Unfortunately, we did not confirm any red candidate at $z = 2.02$ in the surrounding of MRC~0156-252. 
All spectroscopic member galaxies except the radio galaxy have very blue $Y - H$ colours. 
% which are most probably due to on-going star-formation.
% since their spectra do not show any sign of AGN activity. 
Further spectroscopy at redder wavelengths will be necessary to confirm the presence of a red sequence in the galaxy structure.

\begin{figure} 
\begin{center} 
\includegraphics[width=7cm,angle=0]{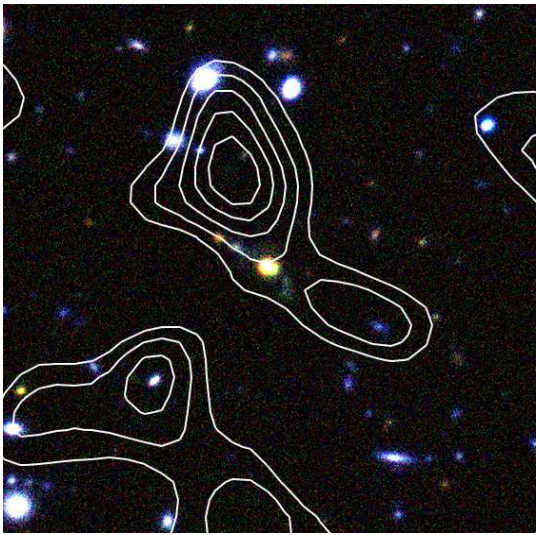}
\end{center}
\caption{Three-colour image of the $1\arcmin \times 1\arcmin$ region around MRC~0156-252 (RGB = $K_sHY$) with
$0.2-6$~keV X-ray `extended emission' contours overlaid in white. See Overzier et al.~(2005) for details.}
\label{vlaxray}
\end{figure}

%\subsection{An X-ray extended emission from the ICM?}
\subsection{X-ray extended emission from the ICM?}

The field of MRC~0156-252 was imaged by {\it Chandra} (Overzier et al.~2005; O05 hereafter) for $20$~ks. 
Along with the study of non-thermal X-ray emission from the radio galaxy core and extended emission from 
the radio lobes, O05 searched for evidence of extended thermal X-ray emission from a potential intracluster 
medium (ICM). MRC~0156-252 was the only field in their sample of five radio galaxies at  $2.0 < z < 2.6$ that 
showed a trace of extended X-ray emission. Figure~\ref{vlaxray} shows the X-ray contours of this extended 
emission --- obtained by smoothing the original $0.2-6$~keV X-ray image by a $10\arcsec$ FWHM Gaussian 
kernel after removing the X-ray counts from the radio galaxy and point sources in the field --- overlaid on a 
3-colour image of the $1\arcmin \times 1\arcmin$ region around MRC~0156-252. 

O05 estimated a tentative $2 \times 10^{-15}$~erg cm$^{-2}$ s$^{-1}$ for the $0.5-2$~keV unabsorbed 
flux for the extended emission. Similar X-ray flux estimations were derived in galaxy proto-clusters at comparable 
high redshifts \citep{Tanaka2010, Papovich2010, Gobat2011}, although, as for the MRC~0156-252 field, both 
these proto-clusters also host a number of X-ray AGN that made it challenging to estimate a reliable X-ray 
flux for the ICM. These estimations are still however $\sim$$5-10$ times fainter than the X-ray emission in the 
highest redshift confirmed X-ray selected galaxy clusters \citep[e.g.,~$z \sim 1.6$;][]{Fassbender2011} 
or mid-infrared selected galaxy clusters \citep[$z=1.75$;][]{Stanford2012} and \citep[$z \sim 1.4-1.5$;][]{Brodwin2011}
which are most probably more advanced collapsed galaxy structures than the present proto-cluster.

No overdensity of red galaxies is visible at the position of the X-ray emission (Fig.~9). Although O05 
also found evidence for extended X-ray emission related to inverse Compton processes in the regions covered by the 
extended radio structure of MRC~0156-252, these regions were excluded entirely when searching for the large-scale 
extended emission shown in Fig.~9. However, due to its closeness to the radio galaxy and its extension to both sides 
of the radio galaxy along the radio axis could indicate that inverse Compton scattering of cosmic microwave background 
(CMB) photons is occurring at much larger scales than those probed by the present radio morphology of MRC~0156-252.
%the trace of extended X-ray emission is significant at a $3\sigma$ level, its closeness to the radio galaxy and its 
%extension to both sides of the radio galaxy along the radio axis could also point towards another origin such as an 
%extended inverse Compton scattering of cosmic microwave background (CMD) photons on the electrons of the radio jet. 
\citet{Finoguenov2010} lists a sample of inverse Compton `ghosts' associated with radio jets whose X-ray fluxes 
are consistent with the X-ray extended emission found in the field of MRC~0156-252. Recent works also suggest 
that the ICM X-ray emission could also come from the far-infrared photons associated from dusty starbursts
\citep[see, e.g.,][]{Smail2013}.

This result unfortunately shows a major limitation of the technique of searching for high-redshift galaxy clusters 
in the vicinity of radio galaxies since it is challenging to disentangle faint extended, thermal emission of a potential 
(proto-)ICM from thermal/non-thermal extended X-ray emission associated with the radio galaxy itself. No further
interpretation can be made at this point. Deeper X-ray observations will be necessary to draw any firm conclusions.

We further note that high-redshift clusters often host AGN \citep[e.g.,][]{Eastman2007, Martini2006, Martini2007, Galametz2009A,
Pentericci2013}, and even without issues such as the inverse Compton scattering of the CMB in the radio jets/lobes of radio-loud AGN, 
disentangling extended ICM emission from nuclear AGN X-ray emission will be a challenge for distant clusters searches 
in low spatial resolution surveys such as eROSITA.

\section{Summary}

We have used optical multi-object spectroscopy with VLT/FORS2 to target $\sim$$80$ high-redshift galaxy 
candidates in the field of the radio galaxy MRC~0156-252 at $z = 2.02$. 

$\bullet$ We have assigned spectroscopic redshifts to $43$ sources, including $32$ galaxies at $z > 1.5$. The 
large majority of the confirmed high-redshift sources are starburst galaxies.

%The FORS2 observations confirmed that the simple $[3.6] - [4.5]$ colour is very efficient at selecting 
%high-redshift galaxies with $\sim$$80$\% of our spectroscopically confirmed galaxies having red IRAC 
%colours. These FORS2 observations 
%were also the first spectroscopic follow-up of $YHK$-selected galaxies and confirmed the reliability of the 
%near-infrared criterion with only one source out of $15$ confirmed b-$YHK$ galaxies having $z_{\rm spec} < 1.5$.

$\bullet$ The spectroscopic follow-up of two close-by companions of the radio galaxy (sources B and D) showed 
large velocity offsets of hundreds of km~s$^{-1}$.% between the individual components themselves. 
They both show obscured AGN signatures. A new $B$-band optical image
shows that MRC~0156-252 has a very clumpy morphology. Higher-resolution imaging or spectroscopy is 
now required to shed light on what processes are precisely occurring in the vicinity of the radio galaxy and 
on the possible origins of the AGN activity (e.g.,~merger-triggered?).

$\bullet$ The spectroscopic follow-up confirmed the presence of a high-redshift galaxy structure at $z = 2.02$ 
associated with MRC~0156-252. Ten members (including sources A, B and D) were found with redshifts 
within $\mid$$z - 2.020$$\mid$ $< 0.015$, corresponding to velocity offsets $< 1500$~km~s$^{-1}$ at $z = 2.02$. 
Although the small number of 
spectroscopically confirmed galaxy members do not permit us to draw firm conclusions regarding the precise 
nature of this galaxy structure, complementary pieces of evidence (i.e.,~analysis of a possible red sequence 
and X-ray data) suggest the galaxy structure associated with MRC~0156-252 is a forming galaxy cluster 
i.e.,~a proto-cluster. In particular, when selecting the bright red near-infrared selected $z > 1.6$ galaxy candidates 
with colour consistent with red sequence models at $z = 2$ (with $z_f = 4-5$), the field was found $\sim3$ times 
denser than control field. The distribution of the high-redshift galaxy candidates in the field shows a secondary peak in the 
South East, $2$~Mpc away from the main clump. One galaxy in this secondary peak was spectroscopically 
confirmed to lie at a redshift consistent with $z = 2.02$ suggesting that the galaxy structure associated with 
MRC~0156-252 is embedded in a larger scale structure.

\bibliographystyle{apj}
\bibliography{apj-jour,biblio}

\begin{thebibliography}{69}
\expandafter\ifx\csname natexlab\endcsname\relax\def\natexlab#1{#1}\fi

\bibitem[{{Bayliss} {et~al.}(2013)}]{Bayliss2013}
{Bayliss}, M.~B. {et~al.} 2013, ArXiv e-prints

\bibitem[{{Beers} {et~al.}(1990){Beers}, {Flynn}, \& {Gebhardt}}]{Beers1990}
{Beers}, T.~C., {Flynn}, K., \& {Gebhardt}, K. 1990, \aj, 100, 32

\bibitem[{{Bertin} \& {Arnouts}(1996)}]{Bertin1996}
{Bertin}, E. \& {Arnouts}, S. 1996, \aaps, 117, 393

\bibitem[{{Brodwin} {et~al.}(2011)}]{Brodwin2011}
{Brodwin}, M. {et~al.} 2011, \apj, 732, 33

\bibitem[{{Castellano} {et~al.}(2007)}]{Castellano2007}
{Castellano}, M. {et~al.} 2007, \apj, 671, 1497

\bibitem[{{De Breuck} {et~al.}(2010)}]{DeBreuck2010}
{De Breuck}, C. {et~al.} 2010, \apj, 725, 36

\bibitem[{{Doherty} {et~al.}(2010)}]{Doherty2010}
{Doherty}, M. {et~al.} 2010, \aap, 509, 83

\bibitem[{{Eales} \& {Rawlings}(1996)}]{Eales1996}
{Eales}, S.~A. \& {Rawlings}, S. 1996, \apj, 460, 68

\bibitem[{{Eastman} {et~al.}(2007)}]{Eastman2007}
{Eastman}, J. {et~al.} 2007, \apjl, 664, L9

\bibitem[{{Eisenhardt} {et~al.}(2008)}]{Eisenhardt2008}
{Eisenhardt}, P.~R.~M. {et~al.} 2008, \apj, 684, 905

\bibitem[{{Fassbender} {et~al.}(2011)}]{Fassbender2011}
{Fassbender}, R. {et~al.} 2011, New Journal of Physics, 13, 125014

\bibitem[{{Fazio} {et~al.}(2004)}]{Fazio2004A}
{Fazio}, G.~G. {et~al.} 2004, \apjs, 154, 10

\bibitem[{{Finoguenov} {et~al.}(2010)}]{Finoguenov2010}
{Finoguenov}, A. {et~al.} 2010, \mnras, 403, 2063

\bibitem[{{Foley} {et~al.}(2011)}]{Foley2011}
{Foley}, R.~J. {et~al.} 2011, \apj, 731, 86

\bibitem[{{Galametz} {et~al.}(2009{\natexlab{a}})}]{Galametz2009B}
{Galametz}, A. {et~al.} 2009{\natexlab{a}}, \aap, 507, 131

\bibitem[{{Galametz} {et~al.}(2009{\natexlab{b}})}]{Galametz2009A}
---. 2009{\natexlab{b}}, \apj, 694, 1309

\bibitem[{{Galametz} {et~al.}(2010{\natexlab{a}})}]{Galametz2010B}
---. 2010{\natexlab{a}}, \aap, 522, A58+

\bibitem[{{Galametz} {et~al.}(2010{\natexlab{b}})}]{Galametz2010A}
---. 2010{\natexlab{b}}, \aap, 516, A101+

\bibitem[{{Galametz} {et~al.}(2012)}]{Galametz2012}
---. 2012, \apj, 749, 169

\bibitem[{{Galametz} {et~al.}(2013)}]{Galametz2013}
---. 2013, \apjs, 206, 10

\bibitem[{{Garilli} {et~al.}(2010)}]{Garilli2010}
{Garilli}, B. {et~al.} 2010, \pasp, 122, 827

\bibitem[{{Gobat} {et~al.}(2011)}]{Gobat2011}
{Gobat}, R. {et~al.} 2011, \aap, 526, A133+

\bibitem[{{Gobat} {et~al.}(2013)}]{Gobat2013}
---. 2013, ArXiv e-prints

\bibitem[{{Hayashi} {et~al.}(2010)}]{Hayashi2010}
{Hayashi}, M. {et~al.} 2010, \mnras, 402, 1980

\bibitem[{{Hopkins} {et~al.}(2008)}]{Hopkins2008}
{Hopkins}, P.~F. {et~al.} 2008, \apjs, 175, 356

\bibitem[{{Jimenez} \& {Verde}(2009)}]{Jimenez2009}
{Jimenez}, R. \& {Verde}, L. 2009, \prd, 80, 127302

\bibitem[{{Kajisawa} {et~al.}(2006)}]{Kajisawa2006}
{Kajisawa}, M. {et~al.} 2006, \mnras, 371, 577

\bibitem[{{Koyama} {et~al.}(2013)}]{Koyama2013}
{Koyama}, Y. {et~al.} 2013, \mnras, 428, 1551

\bibitem[{{Kuiper} {et~al.}(2011)}]{Kuiper2011}
{Kuiper}, E. {et~al.} 2011, \mnras, 415, 2245

\bibitem[{{Kurk} {et~al.}(2009)}]{Kurk2009}
{Kurk}, J. {et~al.} 2009, \aap, 504, 331

\bibitem[{{Kurk} {et~al.}(2004)}]{Kurk2004B}
{Kurk}, J.~D. {et~al.} 2004, \aap, 428, 793

\bibitem[{{Large} {et~al.}(1981)}]{Large1981}
{Large}, M.~I. {et~al.} 1981, \mnras, 194, 693

\bibitem[{{Lidman} {et~al.}(2008)}]{Lidman2008}
{Lidman}, C. {et~al.} 2008, \aap, 489, 981

\bibitem[{{Lotz} {et~al.}(2013)}]{Lotz2013}
{Lotz}, J.~M. {et~al.} 2013, \apj, 773, 154

\bibitem[{{Martini} {et~al.}(2006){Martini}, {Kelson}, {Kim}, {Mulchaey}, \&
  {Athey}}]{Martini2006}
{Martini}, P., {Kelson}, D.~D., {Kim}, E., {Mulchaey}, J.~S., \& {Athey}, A.~A.
  2006, \apj, 644, 116

\bibitem[{{Martini} {et~al.}(2007){Martini}, {Mulchaey}, \&
  {Kelson}}]{Martini2007}
{Martini}, P., {Mulchaey}, J.~S., \& {Kelson}, D.~D. 2007, \apj, 664, 761

\bibitem[{{McCarthy}(1993)}]{McCarthy1993}
{McCarthy}, P.~J. 1993, \pasp, 105, 1051

\bibitem[{{McCarthy} {et~al.}(1990)}]{McCarthy1990}
{McCarthy}, P.~J. {et~al.} 1990, \aj, 100, 1014

\bibitem[{{McCarthy} {et~al.}(1992)}]{McCarthy1992}
---. 1992, \apj, 386, 52

\bibitem[{{Nastasi} {et~al.}(2011)}]{Nastasi2011}
{Nastasi}, A. {et~al.} 2011, \aap, 532, L6

\bibitem[{{Overzier} {et~al.}(2005)}]{Overzier2005}
{Overzier}, R.~A. {et~al.} 2005, \aap, 433, 87

\bibitem[{{Overzier} {et~al.}(2008)}]{Overzier2008}
---. 2008, \apj, 673, 143

\bibitem[{{Papovich}(2008)}]{Papovich2008}
{Papovich}, C. 2008, \apj, 676, 206

\bibitem[{{Papovich} {et~al.}(2010)}]{Papovich2010}
{Papovich}, C. {et~al.} 2010, \apj, 716, 1503

\bibitem[{{Pentericci} {et~al.}(1999)}]{Pentericci1999}
{Pentericci}, L. {et~al.} 1999, \aap, 341, 329

\bibitem[{{Pentericci} {et~al.}(2000)}]{Pentericci2000}
---. 2000, \aap, 361, L25

\bibitem[{{Pentericci} {et~al.}(2001)}]{Pentericci2001}
---. 2001, \apjs, 135, 63

\bibitem[{{Pentericci} {et~al.}(2013)}]{Pentericci2013}
---. 2013, \aap, 552, A111

\bibitem[{{Planck Collaboration}(2013)}]{Planck2013}
{Planck Collaboration}. 2013, arXiv:1303.5080

\bibitem[{{Seymour} {et~al.}(2007)}]{Seymour2007}
{Seymour}, N. {et~al.} 2007, \apjs, 171, 353

\bibitem[{{Smail} \& {Blundell}(2013)}]{Smail2013}
{Smail}, I. \& {Blundell}, K. 2013, ArXiv e-prints

\bibitem[{{Stalder} {et~al.}(2013)}]{Stalder2013}
{Stalder}, B. {et~al.} 2013, \apj, 763, 93

\bibitem[{{Stanford} {et~al.}(2006)}]{Stanford2006}
{Stanford}, S.~A. {et~al.} 2006, \apjl, 646, L13

\bibitem[{{Stanford} {et~al.}(2012)}]{Stanford2012}
---. 2012, \apj, 753, 164

\bibitem[{{Steidel} {et~al.}(2005)}]{Steidel2005}
{Steidel}, C.~C. {et~al.} 2005, \apj, 626, 44

\bibitem[{{Stern} {et~al.}(2005)}]{Stern2005}
{Stern}, D. {et~al.} 2005, \apj, 631, 163

\bibitem[{{Stern} {et~al.}(2012)}]{Stern2012}
---. 2012, \apj, 753, 30

\bibitem[{{Tanaka} {et~al.}(2010){Tanaka}, {Finoguenov}, \&
  {Ueda}}]{Tanaka2010}
{Tanaka}, M., {Finoguenov}, A., \& {Ueda}, Y. 2010, \apjl, 716, L152

\bibitem[{{Tanaka} {et~al.}(2007)}]{Tanaka2007}
{Tanaka}, M. {et~al.} 2007, \mnras, 377, 1206

\bibitem[{{Tanaka} {et~al.}(2013)}]{Tanaka2013}
---. 2013, \apj, 772, 113

\bibitem[{{Tran} {et~al.}(2010)}]{Tran2010}
{Tran}, K.-V.~H. {et~al.} 2010, \apjl, 719, L126

\bibitem[{{Vanderlinde} {et~al.}(2010)}]{Vanderlinde2010}
{Vanderlinde}, K. {et~al.} 2010, \apj, 722, 1180

\bibitem[{{Venemans} {et~al.}(2004)}]{Venemans2004}
{Venemans}, B.~P. {et~al.} 2004, \aap, 424, L17

\bibitem[{{Venemans} {et~al.}(2007)}]{Venemans2007}
---. 2007, \aap, 461, 823

\bibitem[{{Villar-Mart{\'{\i}}n} {et~al.}(2007)}]{Villar2007}
{Villar-Mart{\'{\i}}n}, M. {et~al.} 2007, \mnras, 378, 416

\bibitem[{{Williams} {et~al.}(2009)}]{Williams2009}
{Williams}, R.~J. {et~al.} 2009, \apj, 691, 1879

\bibitem[{{Wuyts} {et~al.}(2009)}]{Wuyts2009}
{Wuyts}, S. {et~al.} 2009, \apj, 706, 885

\bibitem[{{Wylezalek} {et~al.}(2013)}]{Wylezalek2013}
{Wylezalek}, D. {et~al.} 2013, \apj, 769, 79

\bibitem[{{Zirm} {et~al.}(2008)}]{Zirm2008}
{Zirm}, A.~W. {et~al.} 2008, \apj, 680, 224

\end{thebibliography}

\newpage

\begin{appendix}

\section{Spectroscopic redshifts}

\begin{table*}
\caption{Structure members in the redshift range $2 < z < 2.03$}
\centering
\begin{tabular}{l l l l l l l}
ID	&	R.A.$^{\mathrm{a}}$	&	Dec.$^{\mathrm{a}}$	&	Selection	&	$z_{\rm spec}$$^{\mathrm{b}}$	& Absorption / Emission lines$^{\mathrm{c}}$	&	Galaxy type	\\
	&	J2000			&	J2000			&			&							&	or spectral features			&			\\
\hline
4352 		& 01:58:33.48 & -24:59:32.01 		& Comp. A 	&	$2.0256 \pm 0.0002$	&	Ly$\alpha$$^*$; CIV$^*$; HeII$^*$; CIII]$^*$ &	RG	\\			
4351			& 01:58:33.68 & -24:59:30.56 		& Comp. B 		&	$2.0171 \pm 0.0004$ 	& 	Ly$\alpha$$^*$; CIV$^*$; HeII$^*$; CIII]$^*$	&		Type-2 AGN	\\
4365 		& 01:58:33.31 & -24:59:34.48 		& Comp. D 		&	$2.0224 \pm 0.0002$	&	Ly$\alpha$$^*$; CIV$^*$; HeII$^*$; CIII]$^*$	&		Type-2 AGN	\\
\hline
3803 		& 01:58:34.99 & -24:58:36.83 		& IRAC				&	$2.0188 \pm 0.0002$ 	&	SiII$_{1260}$; OI/SiII; CII; SiIV; SiII$_{1526}$; CIV; FeII; AlII	&	star-forming \\
3344 		& 01:58:30.59 & -24:57:52.92 		& IRAC				&	$2.0205 \pm 0.0003$ 	&	SiII$_{1260}$; OI/SiII; CII; SiIV; SiII$_{1526}$; CIV; FeII; AlII; &	\\
			&			&				&					&					&	SiII$_{1808}$; AlIII & star-forming 	\\
662 			& 01:58:40.62 & -25:02:47.57 		& b-$YHK$/IRAC		&	$2.0223 \pm 0.0004$ 	&	CII; SiII$_{1526}$; CIV; HeII; SiII$_{1808}$ & star-forming 	\\
3886 		& 01:58:33.58 & -24:58:53.48 		& IRAC				&	$2.0193 \pm 0.0007$	&	SiII$_{1260}$; OI/SiII; CII; SiIV; CIV; AlII; SiII$_{1808}$ & star-forming 	\\
451 			& 01:58:28.00 & -25:03:10.98 		& b-$YHK$/IRAC		&	$2.0196 \pm 0.001$ (B)	&	OI/SiII; CII; SiII$_{1526}$; CIV; AlII & star-forming 	\\
4030 		& 01:58:30.05 & -24:59:04.44 		& IRAC				&	$2.0194 \pm 0.004$ (B)	&	Ly$\alpha$$^*$; SiII$_{1260}$; CII; CIV & star-forming 	\\
3875 		& 01:58:36.13 & -24:58:44.60 		& b-$YHK$/IRAC		&	$2.0089 \pm 0.003$ (B)	&	SiII$_{1526}$; CIV; AlIII 	&	star-forming	\\
\end{tabular}
\label{spectro}
\begin{list}{}{}
\item[$^{\mathrm{a}}$] Coordinates are derived from the $B$-band pre-imaging.
\item[$^{\mathrm{b}}$] Quality B redshifts are marked by `(B)' 
\item[$^{\mathrm{c}}$] Emission lines are marked by an asterisk.
\end{list}
%\begin{list}{}{}
%\item[$^{\mathrm{a}}$] Coordinates are derived from the $B$-band pre-imaging.
%\item[$^{\mathrm{b}}$] Quality B redshifts are marked by `(B)' 
%\item[$^{\mathrm{c}}$] Emission lines are marked by an asterisk. 
%\end{list}
\end{table*}

\begin{table*}
\caption{Other spectroscopic redshifts}
\centering
\begin{tabular}{l l l l l ll}
ID	&	R.A.$^{\mathrm{a}}$	&	Dec.$^{\mathrm{a}}$	&	Selection	&	$z_{\rm spec}$$^{\mathrm{b}}$	& Absorption / Emission lines$^{\mathrm{c}}$	&	Galaxy type	\\
	&	J2000			&	J2000			&			&							&	or spectral features			&				\\
\hline
641 & 01:58:24.28 & -25:02:47.07 & IRAC			&	$3.0767 \pm 0.0003$	&	Ly$\alpha$$^*$; CIV$^*$												&		\\
1794 & 01:58:30.94 & -24:59:50.50 & r-$YHK$/IRAC	&	$2.6595 \pm 0.0004$	&	CII; OI/SiII; SiII$_{1260}$; SiIV; SiII$_{1526}$; AlII							&	star-forming	\\
4228 & 01:58:33.66 & -24:59:19.82 & b-$YHK$		&	$2.6727 \pm 0.0007$ 	&	OI/SiII; SiII$_{1260}$; CII;  SiII$_{1526}$; CIV								&	star-forming	\\
2024 & 01:58:27.60 & -25:00:12.73 & IRAC			&	$2.5708 \pm 0.0002$ 	&	Ly$\alpha$$^*$; CII; SiII$_{1526}$										&		\\
4314 & 01:58:48.71 & -24:59:26.14 & b-$YHK$/IRAC	&	$2.5282 \pm 0.0003$ 	&	Ly$\alpha$$^*$; OI/SiII; CII; SiII$_{1526}$; CIV								&	star-forming	\\
741 & 01:58:29.55 & -25:02:39.78 & b-$YHK$/IRAC		&	$2.3161 \pm 0.0004$ 	&	Ly$\alpha$ (strong abs.); OI/SiII; CII;	SiII$_{1526}$; AlII; SiII$_{1808}$			&	star-forming	\\
4439 & 01:58:32.09 & -24:59:37.91 & b-$YHK$		&	$2.2896 \pm 0.0002$	& 	OI/SiII; CII; SiII$_{1260}$; SiIV; SiII$_{1526}$; AlII, AlIII 						&	star-forming 	\\
3432 & 01:58:45.44 & -24:58:03.60 & IRAC			&	$2.252 \pm 0.001$	&	OI/SiII; CII; SiIV; CIV														&	star-forming	\\
2376 & 01:58:29.80 & -25:00:44.32 & IRAC			&	$2.2198 \pm 0.0015$ 	&	OI/SiII; SiIV; CIV 													&	star-forming 	\\
2250 & 01:58:36.78 & -25:00:32.39 & b-$YHK$/IRAC	&	$2.211 \pm 0.001$	& 	OI/SiII; CIV; CII															&	star-forming	\\
1240 & 01:58:26.41 & -25:01:54.71 & r-$YHK$			&	$2.2066 \pm 0.0002$	& 	CIV$^*$; HeII$^*$													&	Type-2 AGN	\\
696 & 01:58:34.30 & -25:02:46.55 & r-$YHK$/IRAC		&	$2.1249 \pm 0.0003$ 	&	Ly$\alpha$$^*$; CIV$^*$												&	Type-2 AGN	\\
3043 & 01:58:40.18 & -25:01:37.14 & b-$YHK$/IRAC	&	$1.9814 \pm 0.0005$	&	Ly$\alpha$$^*$; OI/SiII; SiII$_{1526}$; AlII								&	star-forming	\\
316 & 01:58:24.30 & -25:03:22.26 & b-$YHK$/IRAC		&	$1.9663 \pm 0.0005$ 	&	OI/SiII; CII; SiIV; SiII$_{1526}$; CIV; FeII; OIII; AlII							&	star-forming	\\
3812 & 01:58:29.73 & -24:58:34.77 & b-$YHK$		&	$1.9454 \pm 0.0009$	&	OI/SiII; SiII$_{1526}$; CIV; AlII											&	star-forming	\\
1414 & 01:58:29.69 & -25:01:43.90 & IRAC			&	$1.9401 \pm 0.0002$	&	Ly$\alpha$$^*$; SiII$_{526}$; CIV 										&		\\
3622	 & 01:58:33.54 & -24:58:18.21 & IRAC			&	$1.9231 \pm 0.0005$	& 	OI/SiII; CII; SiIV; SiII$_{1526}$; CIV										&	star-forming \\
1011 & 01:58:30.58 & -25:02:13.87 & IRAC			&	$1.732 \pm 0.002$ (B)	&	CIV																&		\\
2358 & 01:58:32.82 & -25:00:42.43 & IRAC			&	$1.6680 \pm 0.0003$	&	CII; SiII$_{1526}$; SiIV												&	star-forming	\\
1901 & 01:58:32.39 & -25:00:00.27 & IRAC			&	$1.672 \pm 0.001$		&	SiIV; CIV; OIII														&	star-forming	\\
4280 & 01:58:39.54	& -24:59:33.21 & IRAC			&	$1.6674 \pm 0.001$ (B)	&	OI/SiII; SiIV; CIV													&	star-forming \\
2225 & 01:58:31.09 & -25:00:30.00 & IRAC			&	$1.6309 \pm 0.0005$	&	OI/SiII; SiIV; SiII$_{1526}$; CIV; AlII; AlIII									&	star-forming	\\
874 & 01:58:40.01 & -25:02:27.34 & b-$YHK$/IRAC		&	$1.5364 \pm 0.0003$  	&	CIV$^*$; HeII$^*$ ; CIII]$^*$; MgII$^*$									&	QSO \\
1019 & 01:58:40.57 & -25:02:14.33 & b-$YHK$/IRAC	&	$1.530 \pm 0.001$	&	CII; SiIV; CIV; AlII													&	star-forming	\\
\hline
168 & 01:58:27.19 & -25:03:36.90 & IRAC			&	$0.7158 \pm 0.0002$	&	[OII]$^*$															&		\\
3850 & 01:58:29.80 & -24:58:44.71 & IRAC			&	$0.4975 \pm 0.0001$	&	[OII]$^*$; Balmer break												&		 \\
3795 & 01:58:41.62 & -24:58:38.19 & IRAC			&	$0.4833 \pm 0.0001$	&	[OII]$^*$; $4000$\AA~break											&	 	\\
268 & 01:58:35.09 & -25:03:17.03 & IRAC			&	$0.4558 \pm 0.0001$ 	&	[NeV]$_{3426}$$^*$; [OII]; Balmer absorption series						&	Type-2 AGN	\\
130 & 01:58:26.31 & -25:03:43.08 & b-$YHK$ 			&	$0.4271 \pm 0.0001$ 	&	[OII]$^*$															&		\\
2025 & 01:58:29.36 & -25:00:11.47 & IRAC			&	$0.4032 \pm 0.0001$ 	&	[OII]$^*$; [NeIII]$_{3967}$$^*$; H$\gamma$$^*$; H$\beta$$^*$; [OIII]$_{5007}$$^*$	&	\\
3642 & 01:58:38.33 & -24:58:17.40 & mid-IR AGN		&	$0.2726 \pm 0.0001$	&	[OII]$^*$; H$\beta$$^*$; [OIII]$_{5007}$$^*$									&		\\
\hline
3890 & 01:58:43.14 & -24:58:51.56 & Alignment		&	$0.4011 \pm 0.0002$	&	[OII]$^*$; E+A 																&		\\
1663 & 01:58:46.63 & -24:57:35.52 & Alignment		&	$0.2856 \pm 0.0002$ 	&	[OII]$^*$; CaHK													&		\\
474 & 01:58:27.03 & -25:02:47.32 	& Alignment 		&	$0.2722 \pm 0.0004$ 	&	CaHK															&		\\
443 & 01:58:35.02 & -25:02:49.86 	& Alignment 		&	$0.2621 \pm 0.0001$ 	&	[OII]$^*$; H$\beta$$^*$; [OIII]$_{5007}$$^*$								&		\\
\end{tabular}		
\label{spectro2}
\begin{list}{}{}
\item[$^{\mathrm{a}}$] Coordinates are derived from the $B$-band pre-imaging.
\item[$^{\mathrm{b}}$] Quality B redshifts are marked by `(B)' 
\item[$^{\mathrm{c}}$] Emission lines are marked by an asterisk.
\end{list}
\end{table*}

\end{appendix}

\end{document}